\documentclass[conference]{IEEEtran}
\IEEEoverridecommandlockouts
\usepackage{cite}
\usepackage{amsmath,amssymb,amsfonts}
\usepackage{algorithmic}
\usepackage{graphicx}
\usepackage{textcomp}
\usepackage{xcolor}
\usepackage{multirow}
\usepackage{booktabs}
\usepackage{url}
\usepackage[normalem]{ulem}
\usepackage{caption}
\useunder{\uline}{\ul}{}
\usepackage{booktabs}
\usepackage{graphicx}
\usepackage{subcaption}
\DeclareRobustCommand*{\IEEEauthorrefmark}[1]{%
    \raisebox{0pt}[0pt][0pt]{\textsuperscript{\footnotesize\ensuremath{#1}}}}

\def\BibTeX{{\rm B\kern-.05em{\sc i\kern-.025em b}\kern-.08em
    T\kern-.1667em\lower.7ex\hbox{E}\kern-.125emX}}

\begin{document}
\title{Burger\includegraphics[scale=0.1]{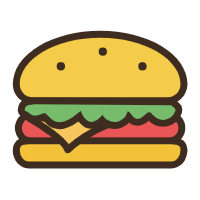}: Robust Graph Denoising-augmentation Fusion and Multi-semantic Modeling for Social Recommendation}

\author{
\IEEEauthorblockN{
Yuqin Lan\IEEEauthorrefmark{1,2},
Weihao Shen\IEEEauthorrefmark{1},
Yuanze Hu\IEEEauthorrefmark{1},
Qingchen Yu\IEEEauthorrefmark{1},
Zhaoxin Fan\IEEEauthorrefmark{1\dagger},
Faguo Wu\IEEEauthorrefmark{1\dagger},
Laurence T. Yang\IEEEauthorrefmark{2\dagger}\thanks{$\dagger$ Zhaoxin Fan , Faguo Wu and Laurence T. Yang are corresponding authors.}
}
\IEEEauthorblockA{
\IEEEauthorrefmark{1}Beijing Advanced Innovation Center for Future Blockchain and Privacy Computing,\\ School of Artificial Intelligence, Beihang University, Beijing, China
}

\IEEEauthorblockA{
\IEEEauthorrefmark{2}School of Cyber Science and Engineering, Huazhong University of Science and Technology, Wuhan, China\\ 
\{lanyq, shenweihao, huyuanze, yuqingchen, zhaoxinf, faguo\}@buaa.edu.cn, ltyang@gmail.com}
}

\maketitle

\begin{abstract}
In the era of rapid development of social media, social recommendation systems as hybrid recommendation systems have been widely applied. Existing methods capture interest similarity between users to filter out interest-irrelevant relations in social networks that inevitably decrease recommendation accuracy, however, limited research has a focus on the mutual influence of semantic information between the social network and the user-item interaction network for further improving social recommendation. To address these issues, we introduce a social \underline{r}ecommendation model with ro\underline{bu}st g\underline{r}aph denoisin\underline{g}-augmentation fusion and multi-s\underline{e}mantic Modeling(Burger). Specifically, we firstly propose to construct a social tensor in order to smooth the training process of the model. Then, a graph convolutional network and a tensor convolutional network are employed to capture user's item preference and social preference, respectively. Considering the different semantic information in the user-item interaction network and the social network, a bi-semantic coordination loss is proposed to model the mutual influence of semantic information. To alleviate the interference of interest-irrelevant relations on multi-semantic modeling, we further use Bayesian posterior probability to mine potential social relations to replace social noise. Finally, the sliding window mechanism is utilized to update the social tensor as the input for the next iteration. Extensive experiments on three real datasets show Burger has a superior performance compared with the state-of-the-art models.
\end{abstract}

\section{Introduction}
With the development of social media, social recommendation as a hybrid recommendation system has attracted extensive attention. Previous research on human social behavior introduced two theories, i.e., the homophily theory \cite{homophily} and the social influence theory \cite{social_influence}. 
The former refers to the fact that people with common interests are more likely to establish social relations, while the latter means that social relations can influence individuals' interests and behaviors.\par
Based on the two theories, the core idea of social recommendation is to mine shared preference patterns among socially-connected users \cite{shared-preference}. Early social recommendations \cite{TrustMF,SocialMF,TrustSVD,PMF,mf_6, yongkang1, yongkang2} are based on social relations, using trust propagation algorithms to predict preferences based on trust degrees, or leveraging friends' interests for collaborative filtering. Later methods \cite{S4Rec,trustgo,design,graphrec} based on Graph Neural Networks (GNNs) construct social networks and user-item networks into graphs, and use graph convolution and graph attention mechanisms for message passing and feature aggregation to capture complex information, accurately mine interests, and improve recommendation quality.\par
\begin{figure}[h]
\centering
\includegraphics[width=0.8\linewidth]{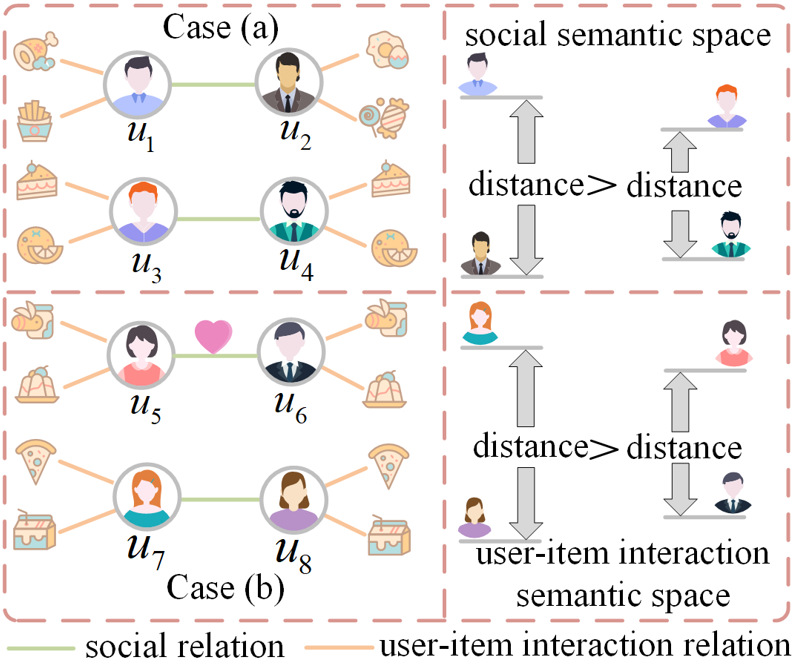}
\caption{A toy example illustrating the mutual influence between the two semantic information.}
\label{introduction}
\end{figure}
Although GNNs-based social recommendations have shown their effectiveness, these methods overlook the fact that interest-irrelevant relations \cite{dunbar} in social networks can hinder performance of the model. On the one hand, users may establish connections with those who don't have the same interests out of social politeness or on a whim. On the other hand, fake accounts and junk information are flooding. These accounts randomly add friends or follow others, which seriously disrupts the normal order of social networks. To alleviate the influence of interest-irrelevant relations, some studies \cite{IDVT, Bottlenecked} attempt to filter out these relations identified based on users' item preferences. 
For example, GDMSR \cite{GDMSR} leverages Transformer \cite{transformer} to capture users' item preferences, calculates the relation confidence between  socially-connected users based on these preferences, and then iteratively deletes the edges of low relation confidence from original social graphs. DSL \cite{DSL} adjusts the self-supervised signals by leveraging the similarity of users' item preferences and filters out misleading social influences.\par
Although these methods are effective in leveraging users' item preferences to identify social noise, we still argue that they suffer from a notable limitation. Specifically, the user-item interaction network and the social network contain different semantic information. These two types of semantic information influence and correlate with each other. Ignoring this mutual influence fails to explicitly preserve and generalize it to socially-connected users, hindering further improvement of the model. Taking Fig.\ref{introduction} as an example, we firstly define two semantic spaces, i.e., the social semantic space and the user-item interaction semantic space. In the case (a), users $u_1$ and $u_2$ are socially related but have minimal shared item preferences. In contrast, users $u_3$ and $u_4$ share both social connections and common item preferences. Consequently, these two semantic differences can result in the distance between $u_3$ and $u_4$ is closer to than that between $u_1$ and $u_2$ in the social semantic space. This case corresponds to the influence of user-item interaction semantic information on social semantic space. Besides, in the case (b), users $u_5$ and $u_6$ have common item preferences and a strong social relation. In contrast, $u_7$ and $u_8$ have common item preferences but a relatively weak social relation. Therefore, these two semantic differences can result in the distance between $u_5$ and $u_6$ is closer to than that between $u_7$ and $u_8$ in the user-item interaction semantic space. This case represents the influence of social semantic information on user-item interaction semantic space. \par
Due to the limitation, we introduce a social \underline{r}ecommendation model with ro\underline{bu}st g\underline{r}aph denoisin\underline{g}-augmentation fusion and multi-s\underline{e}mantic Modeling(Burger), which consists of four main modules: Sliding Window based Social Tensor Construction module, User Dual-preference Joint Mining module, Multi-semantic Modeling module and Graph Denoising-augmentation Fusion module. Firstly, the social graph is converted into a social tensor. Then, a graph convolutional network and a tensor convolutional network are employed respectively to mine users' item preferences and social preferences, and the interest similarity and social similarity are calculated. Next, a bi-semantic coordination loss is proposed, combined with the BPR loss, and the total loss is obtained through weighted summation for multi-semantic modeling. To mitigate the interference of social noise on multi-semantic modeling, a method based on Bayesian posterior probability is employed to identify potential social relations, replace social noise, and construct a new social graph. Finally, the sliding window mechanism is utilized to add the new social graph to the social tensor as the input for the next iteration. We conduct extensive experiments on three real datasets, Ciao, Douban and Yelp. The experimental results validate that our proposed Burger outperforms the state-of-art models.\par
The main contributions of our work can be summarized as follows:
\begin{itemize}
    \item We propose a sliding window mechanism to iteratively construct a social tensor in order to smooth the training process of the model.
    \item To the best of our knowledge, we propose a bi-semantic coordination loss to model the mutual influence of multi-semantic behaviors, which improve model generalization by coordinating relative distances between users in two semantic spaces.
    \item We use Bayesian posterior probability to mine potential social relations to mitigate the interference of social noise on multi-semantic modeling. 
    \item We conduct extensive experiments on three real-world datasets, demonstrating that our model surpasses the state-of-art models.
\end{itemize}

\section{Related Work}
In this section, we introduce the relevant fields about social recommendation based on GNNs and noise-resistant social recommendation.
\subsection{Social Recommendation based on GNNs} Early social recommendation methods \cite{SocialMF,SoReg,TrustSVD,TrustMF} rely on matrix factorization, which fails to deeply capture user social-interest and item-interest connections. Graph neural networks address this by capturing higher-order, robust connections between users and items. For instance, DiffNet \cite{Diffnet} and DiffNet++ \cite{diffnet++} model social network topology, user interests, and social influence. Graphrec \cite{graphrec} uses graph attention networks to integrate user-item and social graph information. FBNE \cite{FBNE} captures high-order implicit relations, but these models are often biased by social noise.
\subsection{Noise-resistant Social Recommendation}
Noise-resistant social recommendation typically involves edge manipulation to clean noisy relations. These methods evaluate the connections between users by mining the preferences on the user-item bipartite graphs. For instance, existing works \cite{probabilistic_1,probabilistic_2} use probabilistic neighbor sampling to learn reliable neighboring nodes adaptively. GDMSR \cite{GDMSR} employs sparsification for a cleaner social graph, while ESRF \cite{ESRF} utilizes GANs to generate denser social connections. On the other hand, some self-supervised models can also resist the interference of noise. They can extract more robust features by mining the self-supervised signals in the data, such as utilizing the contextual information of social networks, time series information, etc., thus effectively reducing the impact of noise on the recommendation results. For instance, Re4 \cite{Re4} and CML \cite{CML} leverage mutual interests to address noise, while DSL \cite{DSL} aligns social and user-item network information through user sampling. Zhou et al. \cite{s3_rec} suggest enhancing sequential recommendation via self-supervised pre-training. MADM\cite{madm} alleviates social noise by maximizing the mutual information between users' social representations and users' item preference representations. \par
Although these methods are effective in resisting the interference of noise. However, some of them ignore the mutual influence of semantic information between the social network and the user-item interaction network, hindering
further improvement of the model.
\section{Methodology}
\begin{figure*}[h]
\centering
\includegraphics[width=1\textwidth]{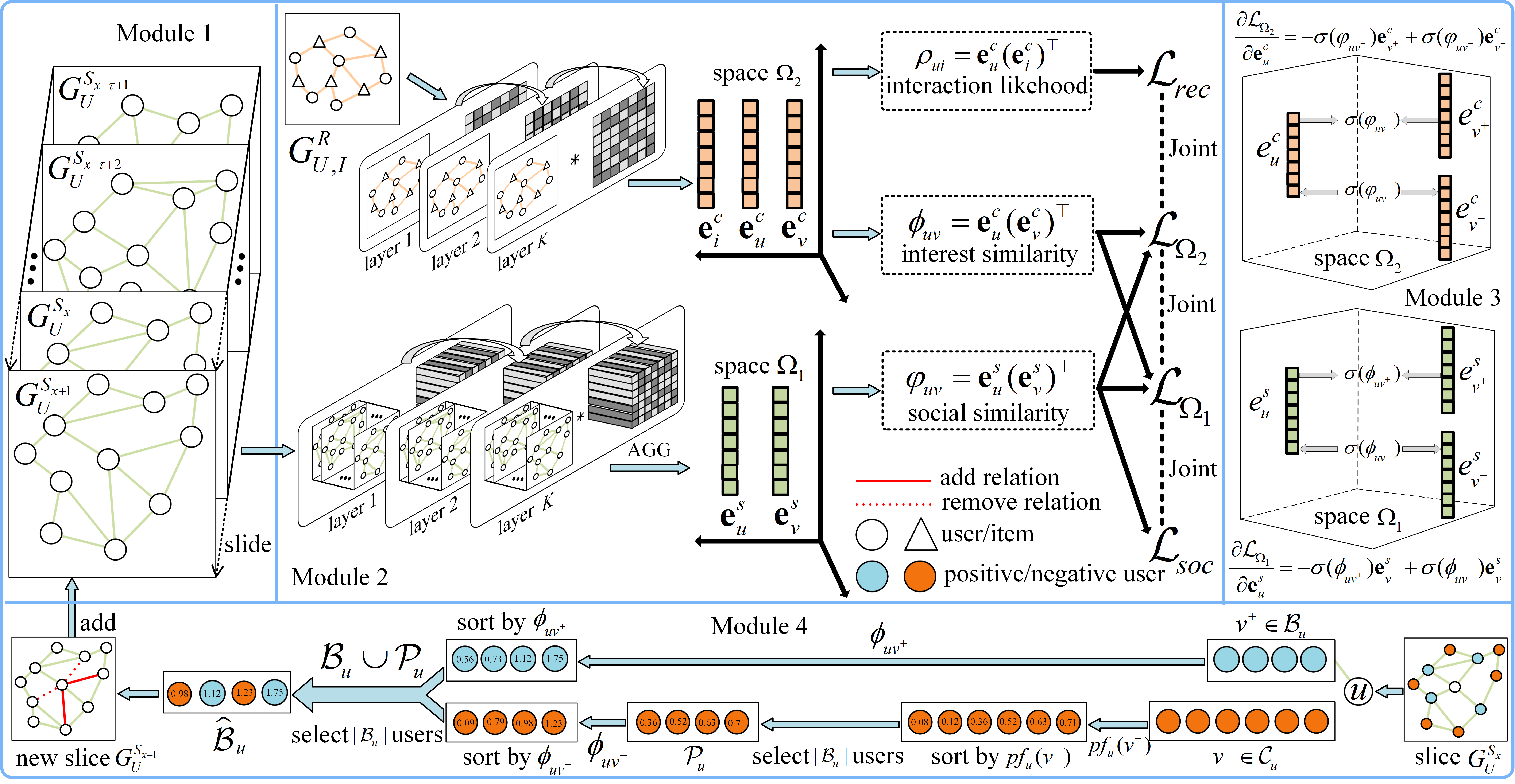}
\caption{The overall framework of Burger model and the process of the (x+1)-th iteration. The Burger model consists of four modules. Module 1 represents sliding window based
social tensor construction module, Module 2 represents user dual-preference joint mining module, Module 3 represents multi-semantic modeling module and Module 4 represents graph denoising-augmentation fusion module.}
\label{Burger Model}
\end{figure*}
The architecture of Burger is shown in Fig.\ref{Burger Model}. The model consists of four main modules, i.e., sliding window based social tensor construction module, user dual-preference joint mining module, multi-semantic modeling module and graph denoising-augmentation fusion module.
\subsection{Notations and Preliminaries}
The Burger model is built upon the foundation of social recommendation
field, where both social network and user-item interaction network are integral components. We define the user set $U=\left\{u_1, u_2, \cdots, u_m\right \} $ and the item set $I = \left \{i_1,i_2,\cdots, i_n \right \}$, where $m$ and $n$ denote the sizes of the respective sets. User-item interaction matrix is represented through $R=\left \{r_{ui}\right \}_{m\times n}$, and social matrix is represented through $S= \left \{s_{uv} \right\}_{m\times m}$. If $r_{ui}=1$, it signifies an interaction between user $u$ and item $i$, otherwise, $r_{ui}=0$. Similarly, $s_{uv}=1$ indicates a explicit social relation between user $u$ and user $v$, while $s_{uv}=0$ signifies no explicit relation. Based on the user-item interaction matrix $R$, a user-item interaction graph $G^{R}_{U,I}=\left \{U,I,E_{U,I}^{R}\right \}$ is constructed, where $E_{U,I}^{R}$ represents a set of interaction edges. Based on the social matrix $S$, the social graph is denoted as $G^{S}_{U}=\left \{U,E_{U}^{S} \right \}$, where $E_{U}^{S}$ represents a set of original social edges between users.\par
According to the above definition, the tasks of the Burger model are as follows: given user-item interaction graph $G^{R}_{U,I}$ and user social graph $G^{S}_{U}$, the goal of Burger is to recommend unobserved items to target users based on the relevance of their embeddings extracted from these graphs.
\subsection{Sliding Window based Social Tensor Construction}
As illustrated in Fig.\ref{Burger Model}, Burger generates a new enhanced social graph in each iteration. To smooth the training process of the model, we propose to use tensor to be the input of the model. Before the first iteration, the operation of randomly adding or deleting edges is used to enhance the social graph due to its simplicity. Given social graph $G^{S}_{U}$, $\tau-1$ enhanced graphs are generated by using random operations, where $\tau$ is a hyperparameter. The $\tau-1$ enhanced graphs along with the original social graph, also called $\tau$ different slices, are denoted as $G^{S_{-\tau+1}}_{U}, G^{S_{-\tau+2}}_{U},\cdots,G^{S_{0}}_{U}$, respectively. These slices are concatenated into a social tensor $\mathcal{T}\in \mathbb{R}^{m\times m \times \tau}$ as follows:
\begin{equation}
    \label{initial tensor}
    \mathcal{T} =concatenate(G^{S_{-\tau+1}}_{U}, G^{S_{-\tau+2}}_{U},\cdots,G^{S_{0}}_{U}).
\end{equation}
After the first iteration, Burger generates an enhanced slice $G^{S_{1}}_{U}$ based on the latest slice $G^{S_{0}}_{U}$ in the social tensor $\mathcal{T}$. The slice $G^{S_{1}}_{U}$ contains less noise and richer information. Inspired by \cite{slide_window}, we propose to use sliding window mechanism \cite{slide_window} to update $\mathcal{T}$. The slice $G^{S_{1}}_{U}$ will be added to $\mathcal{T}$, while the oldest slice $G^{S_{-\tau+1}}_{U}$ in $\mathcal{T}$ will be removed. And the social tensor $\mathcal{T}$ is updated as follows:
\begin{equation}
    \mathcal{T} =concatenate(G^{S_{-\tau+2}}_{U}, G^{S_{-\tau+3}}_{U},\cdots,G^{S_{1}}_{U}).
\end{equation}
The process of subsequent iterations is similar to that of the first iteration. For example, after the (x+1)-th iteration, Burger generates an enhanced slice $G^{S_{x+1}}_{U}$ based on the slice $G^{S_{x}}_{U}$. And the social tensor $\mathcal{T}$ is updated as follows:
\begin{equation}
    \mathcal{T} =concatenate(G^{S_{x-\tau+2}}_{U}, G^{S_{x-\tau+3}}_{U},\cdots,G^{S_{x+1}}_{U}).
\end{equation}
\subsection{User Dual-preference Joint Mining}
User Dual-preference Joint Mining module is a key component used to process user preference information. This module aims to simultaneously mine two different types of preferences of users and obtain a more comprehensive and accurate user profile.
\subsubsection{\textbf{users' item preference mining}}
Existing users' item preference modeling largely relies on collaborative filtering methods. In these methods, graph convolutional networks (GCN) excel at capturing collaborative signals on the graph $G^{R}_{U,I}$, leading to improved recommendation performance. Taking the classic GCN model LightGCN\cite{lightGCN} as an example, $\mathbf{E}_{U}\in\mathbb{R}^{m\times d}$ and $\mathbf{E}_{I}\in \mathbb{R}^{n\times d}$ represent the initial embedding matrices of users and items, respectively. The computation process of the $k$-th layer graph convolution in LightGCN is as follows:
\begin{align}
\mathbf{e}_{u}^{(k+1)}=\sum_{i\in\mathcal{N}_u}\frac1{\sqrt{|\mathcal{N}_u|}\sqrt{|\mathcal{N}_i|}}\mathbf{e}_i^{(k)}, \\
\mathbf{e}_{i}^{(k+1)}=\sum_{u\in\mathcal{N}_i}\frac1{\sqrt{|\mathcal{N}_i|}\sqrt{|\mathcal{N}_u|}}\mathbf{e}_u^{(k)},
\end{align}
where $d$ represents the embedding dimension, and $\mathcal{N}_u$ and $\mathcal{N}_i$ represent the one-hop neighbor sets of user $u$ and item $i$ respectively. Finally, using average pooling operation for the $K$ layers embeddings as the embeddings of user and item as follows:
\begin{align}
    \mathbf{e} _{u}^{c}=\frac{1}{K+1}\sum_{k=0}^{K}\mathbf{e} _{u}^{k},\ \  \mathbf{e} _{i}^{c}=\frac{1}{K+1}\sum_{k=0}^{K}\mathbf{e} _{i}^{k},
\end{align}
where $\mathbf{e} _{u}^{c}$ and $\mathbf{e} _{i}^{c}$ represent user $u$'s item preference and item $i$'s attractiveness, respectively.
Naturally, the interest similarity $\phi _{uv}$ between user $u$ and user $v$ can be calculated as follows:
\begin{align}
    \phi _{uv}= \mathbf{e} _{u}^{c}(\mathbf{e} _{v}^{c})^{\top}.
\end{align}
\subsubsection{\textbf{users' social preference mining}}
To extract users' social preference, a Tensor Convolutional Network is employed on tensor $\mathcal{T}$. Given the user's initial embedding matrix $\mathbf{E}_{U}\in\mathbb{R}^{m\times d}$, it is copied into a tensor $\mathcal{E} _{U}\in\mathbb{R}^{m\times d\times \tau }$. And the convolution from the $k$-th to the $(k+1)$-th layer of Tensor Convolutional Network can be described as follows:
\begin{align}
    \mathcal{E} _{U}^{k+1}= \mathcal{T} \ast \mathcal{E} _{U}^{k}.
\end{align}
The convolution results of different layers are averaged to serve as the final user embedding tensor $\mathcal{E} _{U}^{s}$ as follows:
\begin{align}
    \mathcal{E} _{U}^{s}=\frac{1}{K+1}\sum_{k=0}^{K}  \mathcal{E} _{U}^{k}.
\end{align}
It should be noted that $\mathcal{E} _{U}^{s}$ contains $\tau $ slices, which need to be aggregated to obtain the final user social embedding:
\begin{align}
    \mathbf{E}_{U}^{s}  =AGG(\mathbf{E} _{U}^{s,(1)}, \cdots ,\mathbf{E} _{U}^{s,(\tau )}),\ \ \ \mathbf{E} _{U}^{s,(\cdot)} \in \mathcal{E} _{U}^{s}.
\end{align}
where AGG operation can be multilayer perceptron or other methods such as average pooling operation. Naturally, the social similarity $\varphi _{uv}$ between user $u$ and user $v$ can be calculated as follows:
\begin{align}
    \varphi_{uv}= \mathbf{e}_{u}^{s}(\mathbf{e}_{v}^{s})^{\top}, \ \ \mathbf{e}_{\cdot }^{s}\in \mathbf{E} _{U}^{s},
\end{align}
where $\mathbf{e}_{u}^{s}$ represents user $u$'s social preference.
\subsection{Multi-semantic Modeling}
After obtaining the interest similarity and social similarity of users, an intuitive idea is to leverage the recommendation loss to achieve better performance and model the mutual influence of users' semantic behaviors. In most existing contrastive learning approaches \cite{multual-info-max}, they maximize the agreement between positive samples while pushing negative pairs away in the embedding space, as shown below with derived gradients.
\begin{align}
\label{infomax}
    \frac{\partial L_{Infomax}}{\partial{\mathbf{e}}_u}=-{\mathbf{e}}_{v^+}+\sum_{v^{-}}{\mathbf{e}}_{v^-}\frac{\exp{\mathbf{e}}_u^\top{\mathbf{e}}_{v^-}}{\sum_{v^{-}}\exp{\mathbf{e}}_u^\top{\mathbf{e}}_{v^-}},
\end{align}
where $v^{+}$ and $v^{-}$ represent sampled positive users and sampled negative users of anchor user $u$ respectively. For simplicity, we have omitted the vector normalization and the temperature coefficient. And the coefficients in front of the positive and negative user embeddings represent the magnitude, while the signs indicate the direction. Inspired by the Eq.\ref{infomax}, it is necessary to propose a bi-semantic coordination loss to effectively model the mutual influence of semantic information between social graph $G^{S}_{U}$ and user-item interaction graph $G^{R}_{U,I}$. Specifically, we firstly define social semantic space as $\Omega_{1}$ and user-item interaction semantic space as $\Omega_{2}$. This loss enables the interest similarity between users to coordinate their relative positions in the space $\Omega_{1}$, while the social similarity between users coordinates their relative positions in the space $\Omega_{2}$. In the semantic space $\Omega_1$ and $\Omega_2$, the lower the interest/social similarity between the anchor and the positive user, the weaker the force to pull the positive user closer to the anchor, and vice versa. Similarly, the lower the interest/social similarity between the anchor and the negative user, the weaker the force to push the negative user away from the anchor, and vice versa. Since $\phi _{u\cdot}\in\left \{ -\infty,+\infty   \right \}$ and $\varphi_{u\cdot}\in\left \{ -\infty,+\infty   \right \}$, the sigmoid function maps the interest similarity $\phi_{u\cdot}$ and the social similarity to $\sigma(\phi_{u\cdot}) \in (0,1)$ and $\sigma(\varphi_{u\cdot}) \in (0,1)$, which represent the interest magnitude and social magnitude respectively. Therefore, the partial derivative of the loss $\mathcal{L}_{\Omega_{1}}$ which coordinates relative positions in the semantic space $\Omega_{1}$ with respect to the social preference $\mathbf{e} _{u}^{s }$ of anchor $u$ is as follows:
\begin{align}
\label{dao1}
    \frac{\partial\mathcal{L}_{\Omega_1}}{\partial\mathbf{e} _{u}^{s }}=
-\sigma(\phi _{uv^{+}})\mathbf{e} _{v^{+}}^{s}+\sigma(\phi _{uv^{-}})\mathbf{e} _{v^{-}}^{s}.
\end{align}
The partial derivative of the loss $\mathcal{L}_{\Omega_2}$ which coordinates relative positions in the semantic space $\Omega_{2}$ with respect to the item preference $\mathbf{e} _{u}^{c}$ of the anchor user $u$ is as follows:
\begin{align}
\label{dao2}
    \frac{\partial\mathcal{L}_{\Omega_2}}{\partial\mathbf{e} _{u}^{c}}=
-\sigma(\varphi _{uv^{+}})\mathbf{e} _{v^{+}}^{c}+\sigma(\varphi _{uv^{-}})\mathbf{e} _{v^{-}}^{c}.
\end{align}
Based on Eq.\ref{dao1} and Eq.\ref{dao2}, the indefinite partial integration is performed as follows:
\begin{align}
    \int \frac{\partial\mathcal{L}_{\Omega_1}}{\partial\mathbf{e} _{u}^{s }}d\mathbf{e} _{u}^{s }&=\int
\left [   
-\sigma(\phi _{uv^{+}})\mathbf{e} _{v^{+}}^{s}+\sigma(\phi _{uv^{-}})\mathbf{e} _{v^{-}}^{s}\right]d\mathbf{e} _{u}^{s } \nonumber \\
&=-\sigma(\phi _{uv^{+}})\varphi_{uv^{+}}+\sigma(\phi _{uv^{-}})\varphi_{uv^{-}}+C_{1},
\end{align}
\begin{align}
    \int \frac{\partial\mathcal{L}_{\Omega_2}}{\partial\mathbf{e} _{u}^{c}}d\mathbf{e} _{u}^{c}&=\int
\left [   
-\sigma(\varphi _{uv^{+}})\mathbf{e} _{v^{+}}^{c}+\sigma(\varphi _{uv^{-}})\mathbf{e} _{v^{-}}^{c}\right]d\mathbf{e} _{u}^{c} \nonumber \\
&=-\sigma(\varphi _{uv^{+}})\phi_{uv^{+}}+\sigma(\varphi _{uv^{-}})\phi_{uv^{-}}+C_{2},
\end{align}
where $C_{1}$ and $C_{2}$ represent constant. 
To minimize the loss $\mathcal{L}_{\Omega_1}$ and $\mathcal{L}_{\Omega_2}$, a rectification operation is applied as follows:
\begin{align}
    \mathcal{L}_{\Omega_1}=\sum_{(u,v^{+},v^{-})}\max\left \{ 0, C_{1}-\sigma(\phi _{uv^{+}})\varphi_{uv^{+}}+\sigma(\phi _{uv^{-}})\varphi_{uv^{-}} \right \}
\end{align}
\begin{align}
    \mathcal{L}_{\Omega_2}=\sum_{(u,v^{+},v^{-})}\max\left \{ 0, C_{2}-\sigma(\varphi _{uv^{+}})\phi_{uv^{+}}+\sigma(\varphi _{uv^{-}})\phi_{uv^{-}} \right \}
\end{align}
Finally, $\mathcal{L}_{\Omega_1}$ and $\mathcal{L}_{\Omega_2}$ together make up the bi-semantic coordination loss.\par
On the other hand, to achieve better recommendation performance, a classical BPR loss \cite{bpr_loss} is utilized. Naturally, the likelihood of user $u$ interacting with item $i$ can be predicted as follows:
\begin{align}
    \rho _{ui}= \mathbf{e} _{u}^{c}(\mathbf{e} _{i}^{c})^{\top}.
\end{align}
For a given user $u$ and its corresponding sampled positive item $i^{+}$ and sampled negative item $i^{-}$, the BPR recommendation loss is calculated as follows:
\begin{align}
    \mathcal{L}_{rec}=- \sum_{(u,i^{+},i^{-})} ln\ \sigma (\rho _{ui^{+}}- \rho _{ui^{-}}).
\end{align}
Similarly, for a user $u$ and its corresponding sampled positive user $v^{+}$ and sampled negative user $v^{-}$, the BPR social loss is calculated as follows:
\begin{align}
    \mathcal{L}_{soc}=- \sum_{(u,v^{+},v^{-})} ln\ \sigma (\varphi _{uv^{+}}- \varphi _{uv^{-}}).
\end{align}
To obtain the final loss, a weighted sum operation of $\mathcal{L}_{rec}$, $\mathcal{L}_{soc}$, $\mathcal{L}_{\Omega_1}$ and $\mathcal{L}_{\Omega_2}$ is calculated as follows:
\begin{align}
        \mathcal{L}=\mathcal{L}_{rec}+\lambda_{1}\mathcal{L}_{soc}+
        \alpha \mathcal{L}_{\Omega_1}+\beta\mathcal{L}_{\Omega_2}
        +\lambda_{2}(||\mathbf{E}_{U}||+||\mathbf{E}_{I}||),
\end{align}
where $\lambda_{1}$, $\lambda_{2}$, $\alpha$ and $\beta$ represent weight hyperparameters. 
\subsection{Graph Denoising-augmentation Fusion}
\label{bayesian posterior}
Merely modeling the mutual influence of semantic information is unreliable, especially $\mathcal{L}_{\Omega_2}$, because social noise can cause the user pairs corresponding to the noise to have their distances forcibly coordinated in the semantic space $\Omega_2$, which can lead to poor performance. Therefore, it is necessary to remove social noise. In addition, only focusing on discarding irrelevant social connections makes sparse social networks even sparser, meaning social representations may not be sufficient for recommendation.
Therefore, it is necessary to mine potential friends with the same item preferences, which can improve the generalization ability and performance of the model. Based on the two ideas, this module can be divided into two steps: First, identify potential friends from those not observed by the anchor user; second, identify potential friends who have shared item preferences with the anchor user and replace noisy users. Specifically, for user $u$, the set of $u$'s friends observed in the social graph to be enhanced is denoted as $\mathcal{B}_{u}$, i.e., the set of positive users and $v^+\in \mathcal{B}_{u}$. The set of friends not observed by the user $u$ can then be easily obtained: $\mathcal{C}_{u}  = U-\mathcal{B}_{u}-u$, i.e., the set of negative users. An intuitive idea is to identify potential friends from $\mathcal{C}_{u}$. However, identifying potential friends is non-trivial due to a lack of ground-truth labels. \par
For this problem, we attribute it to negative sample sampling. Specifically, in the unobserved set, one type of negative sample is one that anchor users will not interact with, while another type of negative sample may be one that anchor users will interact with; The latter is a pseudo negative sample, while the former is a true negative sample. Inspired by existing work \cite{bns}, we attempt to use Bayesian posterior probability to calculate the probability of unobserved users as pseudo negative samples, and the set of users with the higher probability is selected as the identified potential friends.
Specifically, for user $v^{-}\in \mathcal{C}_{u}$, the probability $pf_{u}(v^{-})$ of being a potential friend of $u$ can be computed as follows:
\begin{align}
\label{pf}
    pf_{u}(v^{-})=\frac{F(\varphi_{uv^{-}})P_{\mathfrak{f}}(v^{-})}{1-F(\varphi_{uv^{-}})-P_{\mathfrak{f}}(v^{-})+2F(\varphi_{uv^{-}})P_{\mathfrak{f}}(v^{-})},
\end{align}
where $F(\varphi_{uv^{-}})=\frac{|\{w|\forall w\in \mathcal{C}_{u},\varphi _{uw}< \varphi _{uv^{-}} \} |}{\left | \mathcal{C}_{u}  \right |  }$, and $P_{\mathfrak{f}}(v^-)$ represents the prior probability that $v^-$ is a potential friend. The prior probability can be determined using exposure rates and other information. The derivation process of this Eq.\ref{pf} is discussed in section \ref{bayesian method}. \par
After obtaining the posterior probability, the $|\mathcal{B}_{u}|$ users with the highest $pf_{u}(v^-)$ from the set $\mathcal{C}_{u}$ are selected as the potential friends of user $u$. The set of potential friends of user $u$, denoted as $\mathcal{P}_{u}$, satisfies the following constraints:
\begin{align}
\label{find_u}
     \forall v^-\in \mathcal{P}_{u},w\in (\mathcal{C}_{u}-\mathcal{P}_{u}),pf_{u}(v^-)\ge pf_{u}(w),
\end{align}
where $|\mathcal{P}_{u}|=|\mathcal{B}_{u}|$, and $\mathcal{C}_{u}-\mathcal{P}_{u}$ represents the set of potential non-friends of user $u$. To simultaneously mine friends from $\mathcal{P}_{u}$ who have shared item preferences with $u$ and remove friends from $\mathcal{B}_{u}$ who don't have shared item preferences with $u$, first combine $\mathcal{P}_{u}$ with $\mathcal{B}_{u}$ into one set and select the $|\mathcal{B}_{u}|$ users with the highest $\phi _{uv^-}$ from this set. The selected users are denoted as the set $\hat{\mathcal{B}}_{u}$. It is evident that set $\hat{\mathcal{B}}_{u}$ satisfies the following constraints:
\begin{align}
\label{find_final_u}
    \forall v^-\in \hat{\mathcal{B}}_{u}, w\in (\mathcal{B}_{u}\cup \mathcal{P}_{u}-\hat{\mathcal{B}}_{u}),\phi _{uv^-}\ge \phi _{uw},
\end{align}
where $|\hat{\mathcal{B}}_{u}|=|\mathcal{B}_{u}|$. Repeat the above procedure to calculate $\hat{\mathcal{B}}_{u}$ for each user in the social graph, according to Eq.\ref{pf}, Eq.\ref{find_u} and Eq.\ref{find_final_u}. Then, based on all users' $\hat{\mathcal{B}}_{u}$, a new denoising-augmentation fusion social graph is constructed.
\section{Experiments}
To evaluate the performance of the proposed Burger model, we conduct extensive experiments on three real datasets. Specifically, our goal is to answer the six questions: \textbf{RQ1}: Does Burger outperform other state-of-the-art models? \textbf{RQ2}: How do different components affect the performance of Burger? \textbf{RQ3}: How do hyperparameters affect the performance of Burger? \textbf{RQ4}: Is Bayesian posterior probability applicable to mining potential friends in social networks? \textbf{RQ5}: Is the social graph enhanced by Burger considered reliable? \textbf{RQ6}: How about the robustness of Burger?
\subsection{Experimental Settings}
\subsubsection{Datasets}
Extensive experiments are conducted on three real datasets: Ciao, Douban, and Yelp. The data is split into training, test sets with a ratio of 7:3. The statistics of the three datasets are shown in Table \ref{datasets table}. \par
\begin{table}[]
\captionsetup{format=plain,labelsep=newline,justification=centering}
\centering
\caption{Table I: Statistics information of datasets.}
\label{datasets table}
\begin{tabular}{c|c|c|c}
\specialrule{1pt}{0pt}{0pt}
Dataset               & Ciao   & Douban & Yelp   \\ \hline
\#users             & 6759   & 2664   & 21461   \\
\#items             & 100884 & 39237  & 109594   \\
\#interactions      & 271025 & 864509 & 902592 \\
interaction density & 0.04\% & 0.83\% & 0.04\% \\
\#social relations  & 110272 & 34106  & 497206  \\
social density      & 0.24\% & 0.48\% & 0.11\% \\ 
\specialrule{1pt}{0pt}{0pt}
\end{tabular}
\end{table}
\subsubsection{Evaluation Metrics} To assess the performance of Burger model and other baselines, we utilize commonly used metrics, i.e., Hit Ratio (HR@N) and Normalized Discounted Cumulative Gain (NDCG@N). Following \cite{leave-one-out}'s suggestion, a leave-one-out strategy is adopted to compute the metrics, i.e., only one sample is retained for the target user in the test set. Therefore, based on this strategy, it is sufficient for experiments to select N = 1 and N = 3.
\subsubsection{Parameter Settings}
In our experiments, for fairness, in the baseline comparison experiments, the embedding dimensions of all models are set to 512, batch size to 1024, regularization parameter $\lambda_2$ to 0.00001, and early stopping patience to 5. The Adam optimizer is used to train the models, while other baseline model parameters are set according to the optimal parameters from the original research. For Burger, the initial social network random enhancement probability is set to 0.01, and the range of convolutional layers for the two graphs is $\left \{ 1,2,3,4 \right \} $. For loss $\mathcal{L}_{\Omega_1}$ and $\mathcal{L}_{\Omega_2}$, set the constants $C_1$ and $C_2$ to 1. For more details, please refer to section \ref{hyperparameter}.
\subsubsection{Baselines} We compare Burger with ten baselines. These ten baselines are mainly divided into three categories: 1) Matrix factorization-based methods: PMF and TrustMF. 2) GNNs-enhanced social recommendations: EATNN, Diffnet, DGRec, NGCF+. 3) Self-supervised social recommendation methods: MHCN, KCGN, SMIN and DSL. The specifics of these ten baseline models are described as follows: 
\begin{itemize}
    \item PMF \cite{PMF}: It is a probabilistic factorization model that uses matrix factorization to map users and items to a common embedding space.
    \item TrustMF \cite{TrustMF}: It decomposes the social and user-item interaction matrices to map users to two separate latent spaces, enhancing recommendation performance.
    \item EATNN \cite{EATNN}: It's a transfer learning model with adaptive features and an attention network to analyze dynamic user-item interactions.
    \item DiffNet \cite{Diffnet}: This model enhances user embeddings iteratively through an influence diffusion mechanism and captures collaborative signals between users and items via a fusion layer.
    \item DGRec \cite{DGRec}: The model aims to capture evolving user behaviors and social influences by merging a recurrent neural network with a graph attention layer.
    \item NGCF+ \cite{NGCF+}: This method enhances collaborative filtering with GNNs by conducting message propagation across a socially-aware user-item relation graph.
    \item KCGN \cite{KCGN}: Enhancing social recommendations involves merging item knowledge with social influence using a multi-task learning approach.
    \item  MHCN \cite{MHCN}: It conducts hypergraph convolution by mining Triangle motifs in the heterogeneous graph.
    \item SMIN \cite{SMIN}: This model boosts social recommendations by employing metapath-guided heterogeneous graph learning and self-supervised signals via mutual information maximization.
    \item DSL \cite{DSL}: It aligns social information with user-item interaction information adaptively by sampling user node pairs.
\end{itemize}
\begin{table*}[]
\caption{Table II: Performance comparison of three metrics HR@1, HR@3 and NDCG@3 on three datasets for all methods. * indicates the p-values of t-tests between Burger and the best baselines are smaller than 0.05.}
\label{comparison}
\resizebox{1\linewidth}{!}{
\begin{tabular}{c|c|c|c|c|c|c|c|c|c|c|c|c|c}
\specialrule{1pt}{0pt}{0pt}
Dataset                  & Metrics & PMF    & TrustMF & EATNN  & {\color[HTML]{000000} DiffNet} & {\color[HTML]{000000} DGRec} & {\color[HTML]{000000} NGCF+} & {\color[HTML]{000000} KCGN} & {\color[HTML]{000000} MHCN} & {\color[HTML]{000000} SMIN} & {\color[HTML]{000000} DSL} & Burger          & Imp\%   \\ \hline
                         & HR@1    & 0.2063 & 0.2217  & 0.2075 & 0.2847                         & 0.2334                       & 0.2936                       & 0.3027                      & 0.3067                      & 0.3100                      & {\ul 0.3401}               & \textbf{0.3603}* & 5.94\%  \\ \cline{2-14} 
                         & HR@3    & 0.3745 & 0.3928  & 0.3781 & 0.4663                         & 0.4096                       & 0.4817                       & 0.4925                      & 0.5035                      & 0.5084                      & {\ul 0.5152}               & \textbf{0.5366}* & 4.15\%  \\ \cline{2-14} 
\multirow{-3}{*}{Ciao}   & NDCG@3  & 0.3302 & 0.3468  & 0.3311 & 0.4019                         & 0.3611                       & 0.4134                       & 0.4281                      & 0.4239                      & 0.4275                      & {\ul 0.4341}               & \textbf{0.4631}* & 6.68\%  \\ \hline
                         & HR@1    & 0.2472 & 0.2671  & 0.2491 & 0.3202                         & 0.2792                       & 0.3378                       & 0.3599                      & 0.3627                      & {\ul 0.3724}                & 0.3679                     & \textbf{0.4140}* & 11.17\% \\ \cline{2-14} 
                         & HR@3    & 0.4892 & 0.5024  & 0.4901 & 0.5688                         & 0.5162                       & 0.5814                       & 0.5948                      & 0.6175                      & {\ul 0.6326}                & 0.6295                     & \textbf{0.6753}* & 6.75\%  \\ \cline{2-14} 
\multirow{-3}{*}{Douban} & NDCG@3  & 0.3973 & 0.4188  & 0.4018 & 0.4807                         & 0.4291                       & 0.4913                       & 0.5050                      & 0.5144                      & {\ul 0.5207}                & 0.5179                     & \textbf{0.5643}* & 8.37\%  \\ \hline
                         & HR@1    & 0.0938 & 0.1103  & 0.0966 & 0.1615                         & 0.1201                       & 0.1798                       & 0.1979                      & 0.1980                      & 0.1994                      & {\ul 0.2066}               & \textbf{0.2194}* & 6.20\%  \\ \cline{2-14} 
                         & HR@3    & 0.2759 & 0.2981  & 0.2798 & 0.3512                         & 0.3072                       & 0.3645                       & 0.3769                      & 0.3788                      & 0.3897                      & {\ul 0.4027}               & \textbf{0.4166}* & 3.45\%  \\ \cline{2-14} 
\multirow{-3}{*}{Yelp}   & NDCG@3  & 0.1901 & 0.2044  & 0.1934 & 0.2663                         & 0.2115                       & 0.2785                       & 0.2970                      & 0.2994                      & 0.3068                      & {\ul 0.3196}               & \textbf{0.3310}* & 3.57\%  \\ 
\specialrule{1pt}{0pt}{0pt}
\end{tabular}
}
\end{table*}
\subsection{Recommendation Performance}
\subsubsection{Overall Performance(RQ1)}
Table \ref {comparison} presents the results of all baselines compared to Burger. The bold-faced values indicate the best results on three datasets, and the underlined value represents the best result among all baselines. From these results, the following conclusions are drawn:\par 
(1) Burger shows better performance on datasets with different characteristics and sparsity levels. On the Douban dataset, Burger significantly outperforms the second-best SMIN baseline in terms of HR@1, HR@3, and NDCG@3 metrics, with performance boosts of 11.17\%, 6.75\%, and 8.37\% respectively. On two other datasets, compared with the second-best DSL model, Burger also has significant improvements in terms of three metrics. Based on these results, the improvements of Burger can be mainly attributed to two reasons: The mutual influence in the two semantic spaces is modeled, enhancing the generalization ability of the model; By mining potential social relations and removing social noise, it can feed back into multi-semantic modeling and further improve the recommendation accuracy.\par
(2) Baselines based on self-supervised signals outperform GNNs-based baselines by an average of 23.76\%. This indicates that self-supervised signals effectively integrate social networks into user's item preference modeling, whereas naively characterizing the entire social network leads to poor performance. And GNNs-based baselines outperform matrix factorization baselines by an average of 35.95\%, as GNNs capture higher-order, complex nonlinear relations, better fitting user preferences and item characteristics.
\subsubsection{Ablation Study(RQ2)}
To verify the effectiveness of each module in Burger, we conducted ablation studies.  Specifically, we compared Burger with the following variants: w/o s represents the removal of the sliding window based social tensor construction, with a single social graph serving as the model input; w/o $\mathcal{L}_{\Omega_1}$ represents the removal of coordinating relative positions in the semantic space $\Omega_1$; w/o $\mathcal{L}_{\Omega_2}$ represents the removal of coordinating relative positions in the semantic space $\Omega_2$; w/o g represents the removal of graph denoising -augmentation fusion. According to these results from Table \ref{ablation study}, we can draw several conclusions as follows:\par
(1) w/o s shows a decrease of 2.83\%, 1.23\%, and 1.27\% in HR@3, and a decrease of 2.01\%, 0.87\%, and 1.42\% in NDCG@3 across three datasets compared with Burger. The extent of these decreases demonstrates that the tensor composed of multiple enhanced graphs enables more stable model learning, thereby enhancing model performance. \par
(2) w/o $\mathcal{L}_{\Omega_1}$ shows a decrease of 3.17\%, 3.76\%, and 2.28\% in HR@3 across three datasets compared with Burger. The extent of these decreases demonstrates that modeling the influence of users' item preferences on the social semantic space $\Omega_1$ can effectively enhance the generalization ability of the model and improve its performance. \par
(3) w/o $\mathcal{L}_{\Omega_2}$ shows a decrease of 1.73\%, 1.39\%, and 1.46\% in HR@3 across three datasets compared with Burger. The extent of these decreases demonstrates that modeling the influence of users' social preferences on the user-item semantic space $\Omega_2$ can also effectively enhance the generalization ability of the model and improve its performance.\par
(4) w/o g shows an obvious decrease of 6.11\%, 5.72\%, and 2.30\% in HR@3, and a decrease of 7.45\%, 8.36\%, and 3.44\% in NDCG@3 across three datasets. This indicates that directly modeling the mutual influence between semantic spaces without enhancing the original social graph can actually harm the model. Conversely, establishing reliable implicit relations and removing unreliable noisy relations can feed back into multi-semantic modeling, thereby improving the model's performance.
\begin{table}[]
\caption{Table III: Abaltion studies for different components in Burger.}
\label{ablation study}
\resizebox{1\linewidth}{!}{
\begin{tabular}{|c|cc|cc|cc|}
\hline
\multirow{2}{*}{Variants} & \multicolumn{2}{c|}{Ciao}            & \multicolumn{2}{c|}{Douban}          & \multicolumn{2}{c|}{Yelp}            \\ \cline{2-7} 
                        & \multicolumn{1}{c|}{HR@3}   & NDCG@3 & \multicolumn{1}{c|}{HR@3}   & NDCG@3 & \multicolumn{1}{c|}{HR@3}   & NDCG@3 \\ \hline
Burger                  & \multicolumn{1}{c|}{0.5366} & 0.4631 & \multicolumn{1}{c|}{0.6753} & 0.5643 & \multicolumn{1}{c|}{0.4166} & 0.3310 \\ \hline
w/o s                   & \multicolumn{1}{c|}{0.5214} & 0.4538 & \multicolumn{1}{c|}{0.6670} & 0.5594 & \multicolumn{1}{c|}{0.4113} & 0.3263 \\ \hline
w/o $\mathcal{L}_{\Omega_1}$                  & \multicolumn{1}{c|}{0.5196} & 0.4392 & \multicolumn{1}{c|}{0.6499} & 0.5476 & \multicolumn{1}{c|}{0.4071} & 0.3219 \\ \hline
w/o $\mathcal{L}_{\Omega_2}$                  & \multicolumn{1}{c|}{0.5273} & 0.4497 & \multicolumn{1}{c|}{0.6659} & 0.5568 & \multicolumn{1}{c|}{0.4105} & 0.3271 \\ \hline
w/o g                   & \multicolumn{1}{c|}{0.5038} & 0.4286 & \multicolumn{1}{c|}{0.6367} & 0.5171 & \multicolumn{1}{c|}{0.4070} & 0.3196 \\ 
\hline
\end{tabular}
}
\end{table}
\subsubsection{Hyper-parameters Sensitivity(RQ3)}
\label{hyperparameter}

We explored how the key hyperparameters in the Burger model affect the model's performance. including $\alpha$, $\beta$, and sliding window size $\tau$. For brevity, we only present results for $\tau$ on the Douban. \par
\begin{figure}[ht]   
    \begin{subfigure}[b]{0.235\textwidth} 
        \includegraphics[width=\textwidth]{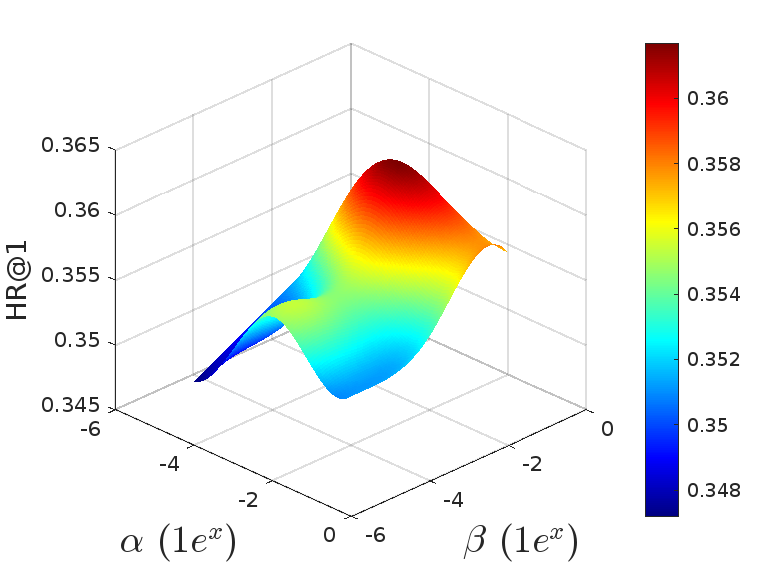}
        \caption{Ciao}
    \end{subfigure}
    \begin{subfigure}[b]{0.235\textwidth} 
        \includegraphics[width=\textwidth]{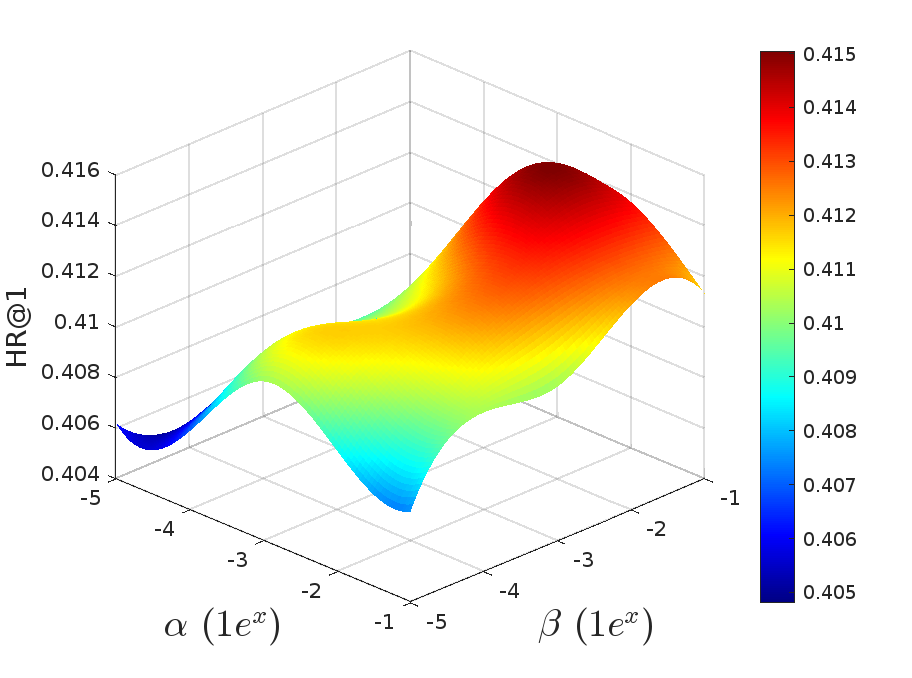}
        \caption{Douban}
    \end{subfigure}
     \begin{subfigure}[b]{0.235\textwidth} 
        \includegraphics[width=\textwidth]{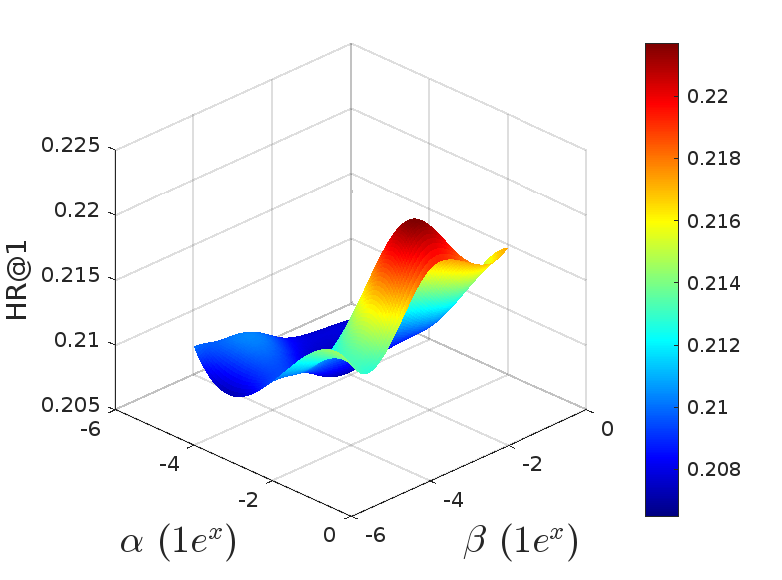}
        \caption{Yelp}
    \end{subfigure}
    \begin{subfigure}[b]{0.235\textwidth} 
        \includegraphics[width=\textwidth]{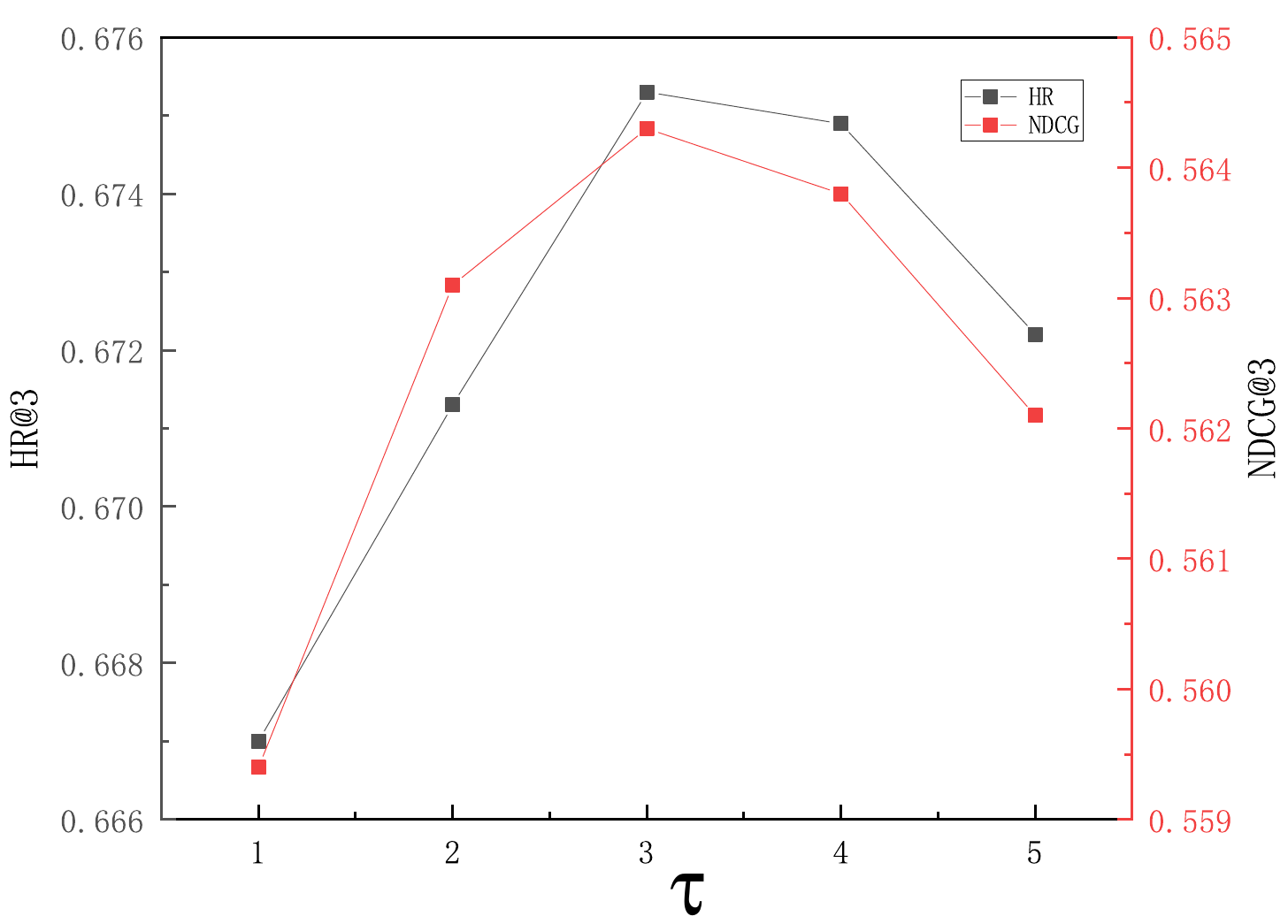}
         \caption{Sensitivity of $\tau$}
    \end{subfigure}
    \caption{Sensitivity of hyperparameter $\alpha$, $\beta$ and sliding window size $\tau$.}
    \label{hyperparameter alpha and beta}
\end{figure}
(1) The weights $\alpha$ and $\beta$ are crucial in balancing the importance of two semantic spaces coordination. The sensitivity of the two parameters is depicted in Fig.\ref{hyperparameter alpha and beta} (a), (b) and (c).  Obviously, on the Ciao, Douban, Yelp, the optimal combinations are found to be $\left \{ \alpha=e^{-3}, \beta=e^{-2} \right \}$, $\left \{ \alpha=e^{-2}, \beta=e^{-2}\right \}$ and $\left \{ \alpha=e^{-2}, \beta=e^{-3}\right \}$ respectively. These optimal values indicate that different datasets have distinct requirements for the balance between the two semantic spaces. For instance, on the Ciao dataset, the optimal combination might suggest that the semantic space related to $\beta$ has a relatively stronger influence on the model's performance, while on the Yelp dataset, the optimal combination implies an opposite emphasis. \par
(2) $\tau$ constrain the sliding window size. The sensitivity is illustrated in Fig.\ref{hyperparameter alpha and beta} (d). Optimal value for Douban is $\tau=3$. Too few tensor slices cause instability, while too many lead to overfitting. Selecting an appropriate sliding window size is of great importance for the stability of the model.
\subsection{Further Analysis}
\subsubsection{\textbf{Bayesian method(RQ4)}}
\label{bayesian method}
The derivation of Bayesian posterior probability requires detailed discussion. Meanwhile, the applicability of Bayesian posterior probability in mining potential friends in social networks needs to be verified.\par
\begin{figure*}[ht]  
    \centering  
    \begin{subfigure}[b]{0.23\textwidth}  
        \includegraphics[width=\linewidth]{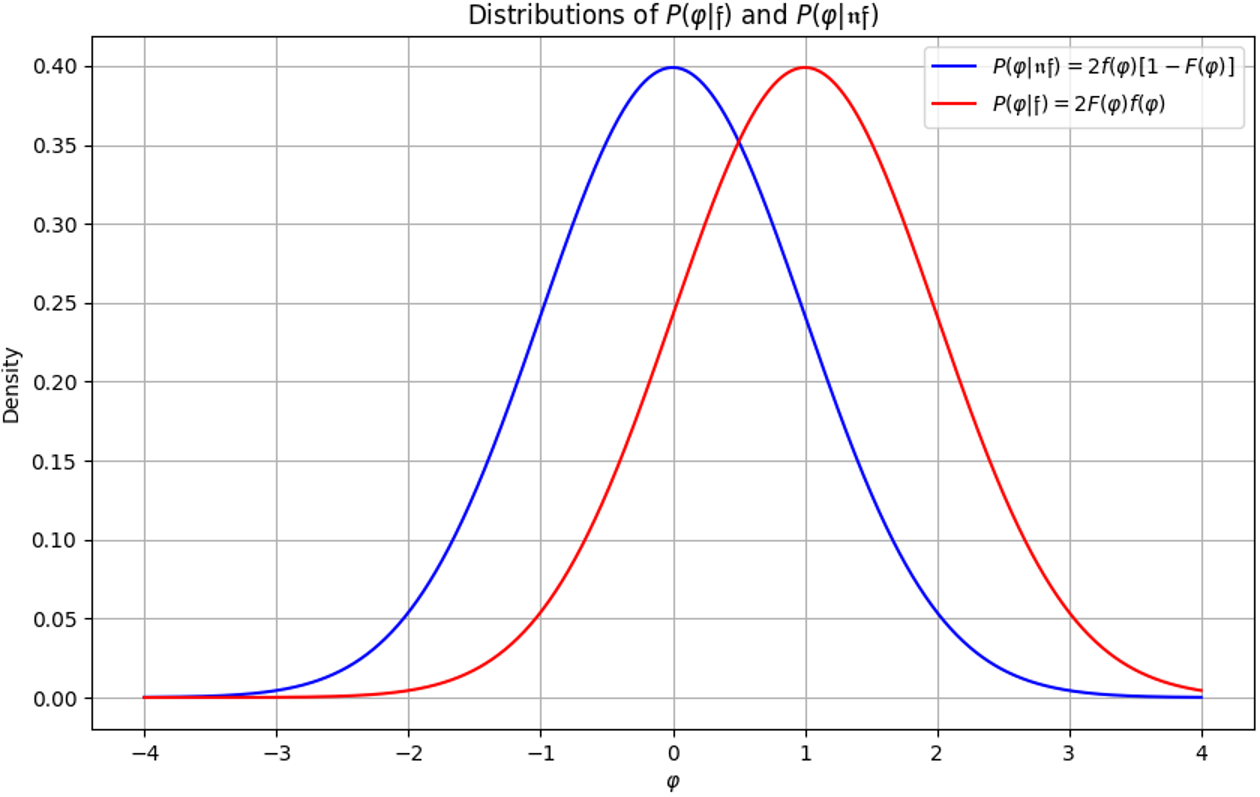}
        \caption{Theoretical distribution}
    \end{subfigure}
    \hfill  
    \begin{subfigure}[b]{0.23\textwidth}
        \includegraphics[width=\linewidth]{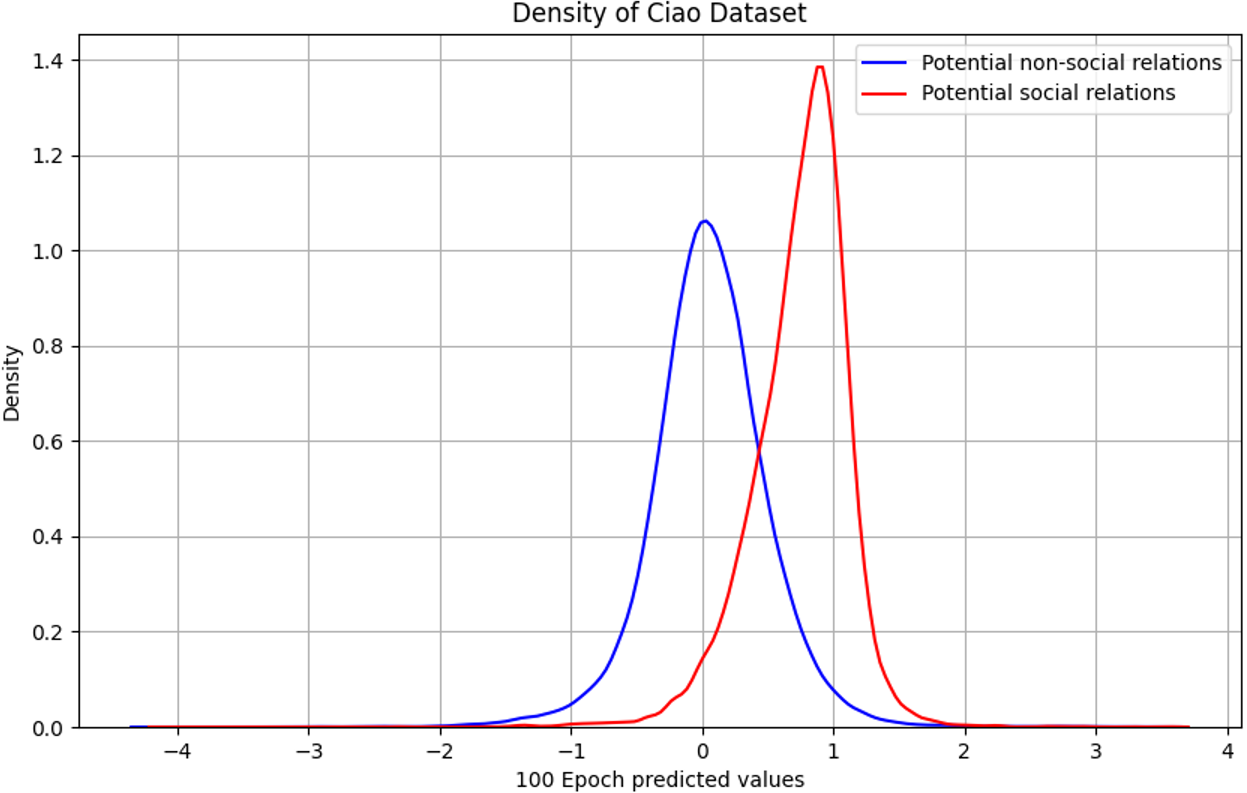}
        \caption{Actual distribution on Ciao}
    \end{subfigure}
    \hfill
    \begin{subfigure}[b]{0.235\textwidth}
        \includegraphics[width=\linewidth]{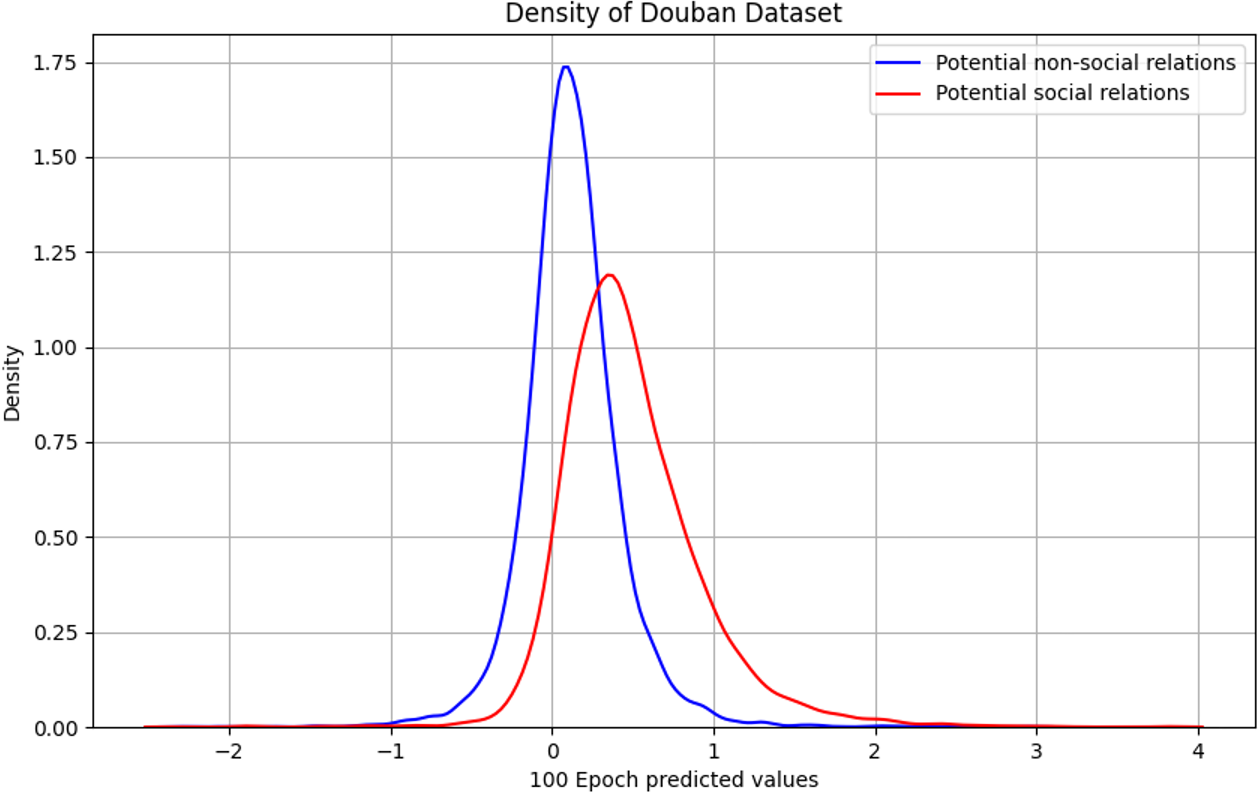}
        \caption{Actual distribution on Douban}
    \end{subfigure}
    \hfill
    \begin{subfigure}[b]{0.23\textwidth}
        \includegraphics[width=\linewidth]{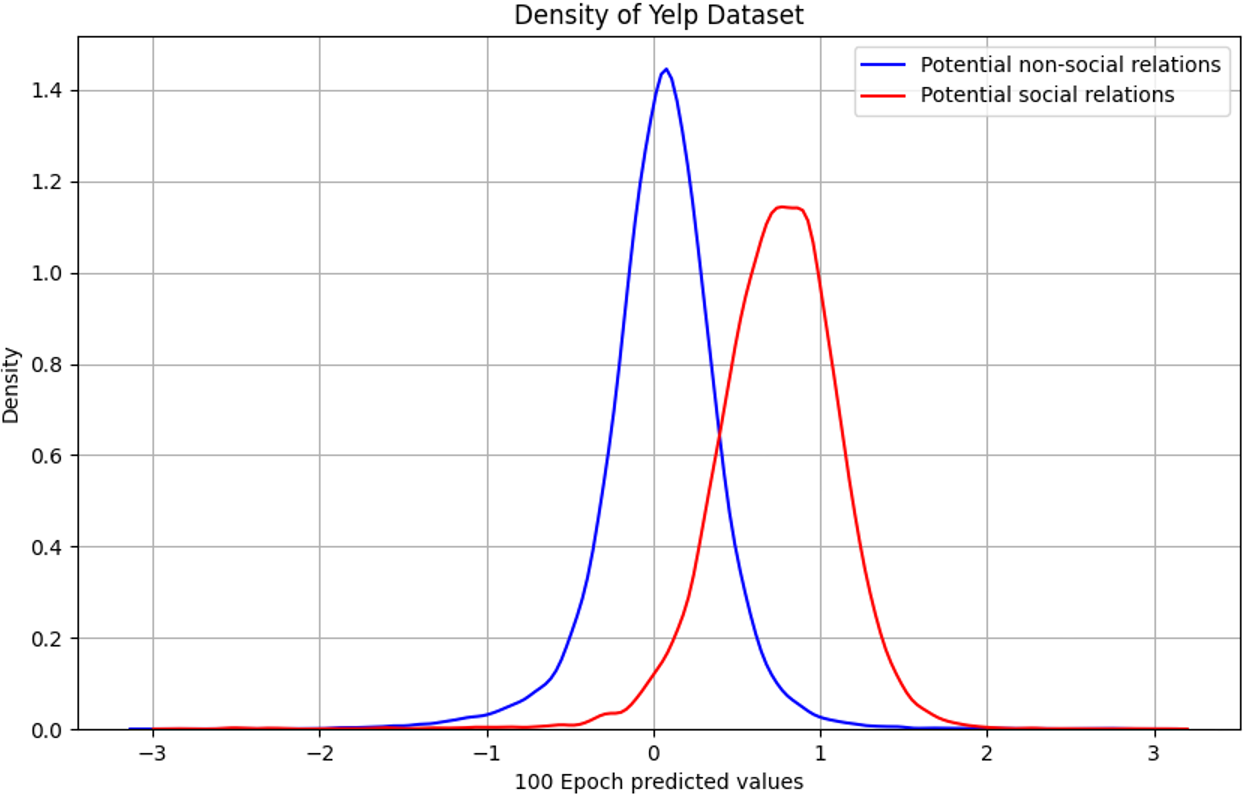}
        \caption{Actual distribution on Yelp}
    \end{subfigure}
    \caption{Theoretical distribution and actual distribution on three datasets.}
    \label{bns distribution}
\end{figure*}

 
Specifically, given user $u$, potential friends of $u$ are denoted as $\mathfrak{f}$ and potential non-friends of $u$ are denoted as $\mathfrak{nf}$. According to the optimization objective of BPR loss($\mathcal{L}_{rec}$ and $\mathcal{L}_{soc}$), the model optimizes to rank positive instances higher than negative instances. This indicates that the ranking often exists, i.e., $\varphi _{u\mathfrak{f} } \ge \varphi _{u\mathfrak{nf}}$, 
where $\varphi _{u\mathfrak{f} }$ and $ \varphi_{u\mathfrak{nf} }$ represent social similarities of potential friends and potential non-friends with $u$, respectively. Next, assume $\varphi$ follows a distribution with distribution function $F(\varphi)$ and density function $f(\varphi)$. Based on the ranking, the conditional densities of $\mathfrak{f}$ and $\mathfrak{nf}$ can be derived as follows:
\begin{align}
    \label{theory distributions 1}
    P(\varphi|\mathfrak{f})&=2F(\varphi)f(\varphi ),\\
    P(\varphi|\mathfrak{nf})&=2f(\varphi)[1-F(\varphi)],
    \label{theory distributions 2}
\end{align}
where $P(\varphi|\mathfrak{f})$ and $P(\varphi|\mathfrak{nf})$ are the theoretical distributions of potential friends and potential non-friends, respectively. 
Given the set $\mathcal{C}_{u}$ and user $v^- \in \mathcal{C}_{u}$, the posterior probability $pf_{u}(v^-)$ can be calculated as: $pf_{u}(v^-)=P(\mathfrak{f} |\varphi _{uv^-})\propto P(\varphi _{uv^-}|\mathfrak{f} )P_{\mathfrak{f}}(v^-)$.
Based on Eq.\ref{theory distributions 1} and Eq.\ref{theory distributions 2}, the Bayesian posterior probability can be naturally derived as follows:
{\footnotesize 
\begin{align*}
    pf_{u}(v^-) &= \frac{P(\mathfrak{f} |\varphi _{uv^-} )}{1} \\
    &= \frac{P(\mathfrak{f} |\varphi _{uv^-} )}{P(\mathfrak{f} |\varphi _{uv^-} )+P(\mathfrak{nf} |\varphi _{uv^-} )} \\
    &\propto \frac{P(\varphi _{uv^-}|\mathfrak{f} )P_{\mathfrak{f}}(v^-)}{
        P(\varphi _{uv^-}|\mathfrak{f} )P_{\mathfrak{f}}(v^-)+P(\varphi _{uv^-}|\mathfrak{nf} )P_{\mathfrak{nf}}(v^-)} \\
    &= \frac{F(\varphi _{uv^-})f(\varphi _{uv^-})P_{\mathfrak{f}}(v^-)}{
        F(\varphi _{uv^-})f(\varphi _{uv^-})P_{\mathfrak{f}}(v^-)+[1-F(\varphi _{uv^-})]f(\varphi _{uv^-})P_{\mathfrak{nf}}(v^-)} \\
    &= \frac{F(\varphi_{uv^-})P_{\mathfrak{f}}(v^-)}{1-F(\varphi_{uv^-})-P_{\mathfrak{f}}(v^-)+2F(\varphi_{uv^-})P_{\mathfrak{f}}(v^-)},
\end{align*}
}
where $P_{\mathfrak{nf}}(v^-)$ represents the prior probability that $v^-$ is a potential non-friend of $u$, and $P_{\mathfrak{f}}(v^-)+P_{\mathfrak{nf}}(v^-)=1$. On the other hand, to verify the applicability of Bayesian posterior probability in mining potential friends in social networks, $F(\varphi)$ is assumed to be a standard normal distribution, and the two theoretical distributions of $ P(\varphi|\mathfrak{f})$ and $P(\varphi|\mathfrak{nf})$ can be plotted as shown in Fig.\ref{bns distribution} (a). Additionally, due to a lack of ground-truth labels in the unobserved data of social networks, we divide the original social network into two parts. One part serves as the observed social relations, another part is regarded as the unobserved potential friends, and the rest are unobserved potential non-friends. Through representation learning of the observed social relations using GNNs, we can obtain the actual probability distributions of potential friends and potential non-friends across three datasets, as shown in Figure \ref{bns distribution} (b), (c), and (d). From these results, it can be observed that the actual distributions align with the theoretical distributions. Therefore, Bayesian posterior probability is applicable to mining potential friends in social networks.
        
       
        
       
\subsubsection{\textbf{The Reliability of Enhanced Social Graph(RQ5)}}
    
        
        
\begin{figure*}[ht]  
    \centering
    \begin{subfigure}[b]{0.23\textwidth}  
        \includegraphics[width=\linewidth]{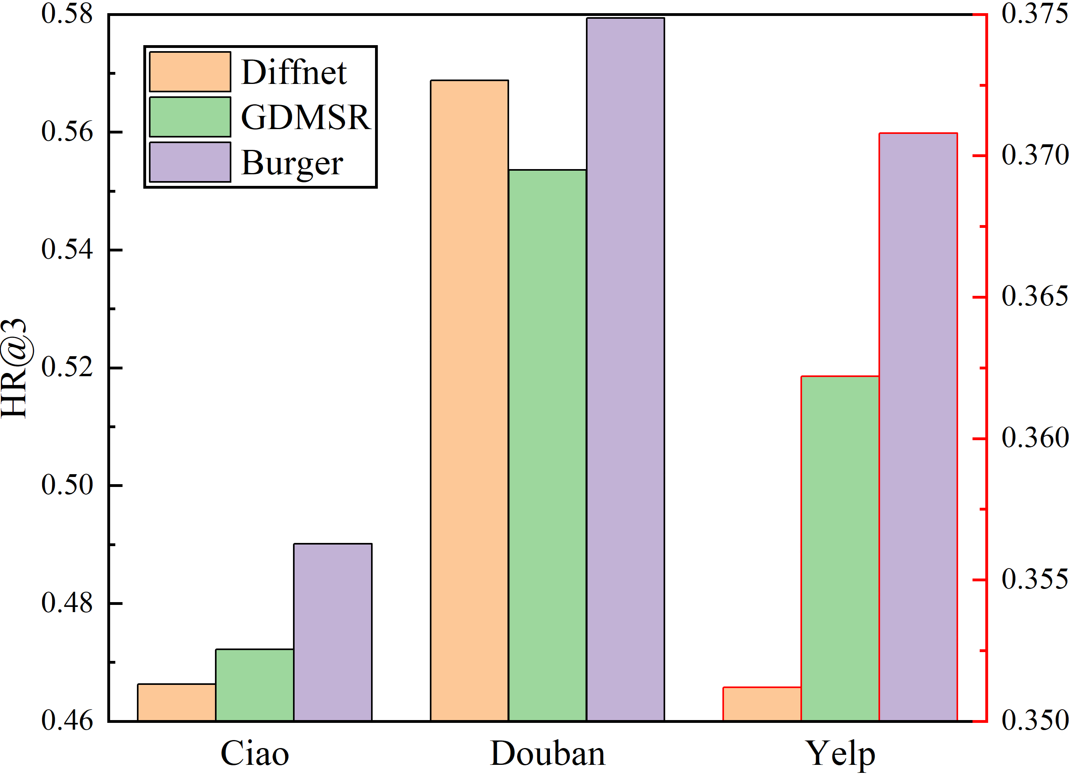}
        \caption{Base Model: DiffNet}
    \end{subfigure}
    \hfill
    \begin{subfigure}[b]{0.23\textwidth}
        \includegraphics[width=\linewidth]{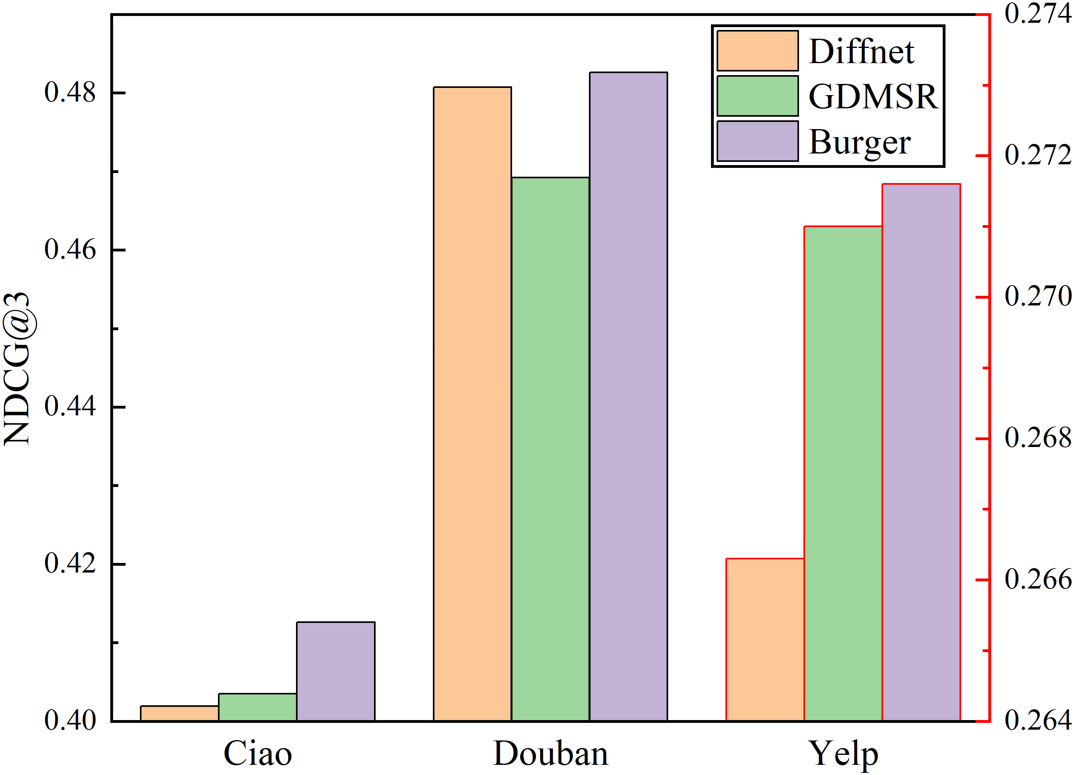}
        \caption{Base Model: DiffNet}
    \end{subfigure}
    \hfill
    \begin{subfigure}[b]{0.23\textwidth}
        \includegraphics[width=\linewidth]{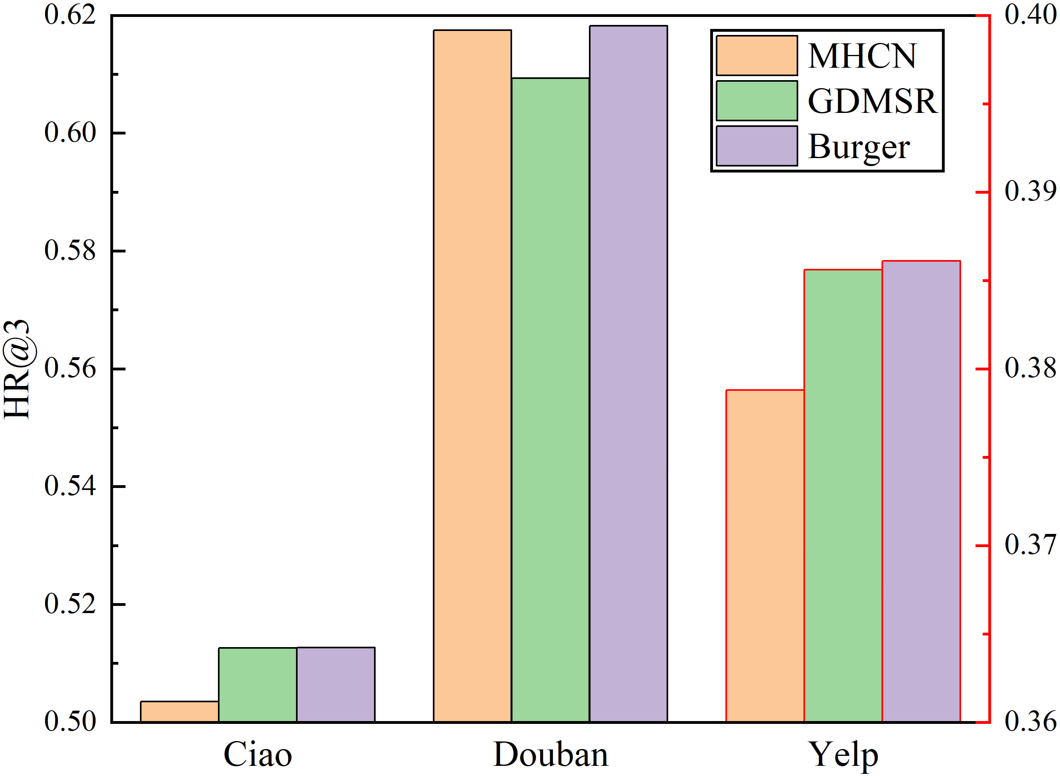}
        \caption{Base Model: MHCN}
    \end{subfigure}
    \hfill
    \begin{subfigure}[b]{0.23\textwidth}
        \includegraphics[width=\linewidth]{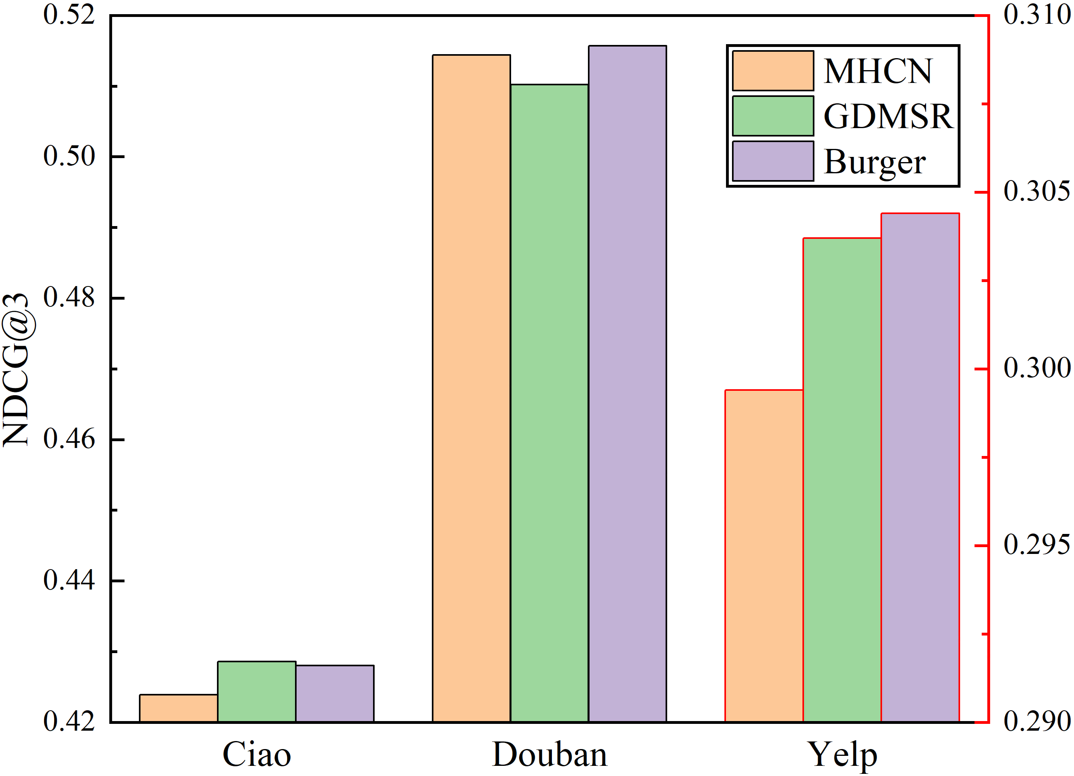}
        \caption{Base Model: MHCN}
    \end{subfigure}
    \caption{The reliability verification of enhanced social graph.}
    \label{fig:denoised_graph_reliable}
\end{figure*}
It's crucial to verify if the enhanced social graph has less noise and more reliable latent relations. We input the Burger-enhanced social graph into Diffnet \cite {Diffnet} and MHCN \cite {MHCN}, using GDMSR \cite {GDMSR} as a baseline. According to the Fig.\ref{fig:denoised_graph_reliable}, we draw the following conclusions: 
(1) The performance of the Burger-enhanced social graph is superior to that of the original social graph. This can be attributed to the fact that the graph denoising-augmentation fusion module can effectively explore latent social relations and reduce social noise. (2) Compared to the baseline, the superiority of our method lies in the fact that mining potential social relations can alleviate the problem of overly sparse social graph after enhancement. The social graph enhanced by the baseline can lead to insufficient learning of user social representations, and our method effectively solves this problem. \par
\subsubsection{\textbf{Robustness Evaluation(RQ6)}}
To evaluate the robustness of Burger, we randomly added a portion of social relations as noise perturbations in the original social network. The synthetic social graph is combined with user-item interaction information to test the performance of Burger. As shown in Table \ref{robust analysis}, compared with the two baselines, Burger's robust performance on the synthetic data lies between them, which also indicates that the robustness of Burger is at a relatively considerable level. This can be attributed to the fact that although the synthetic dataset contains a large amount of noise, the graph denoising-augmentation fusion module can still capture this noise.
\begin{table}[]
\caption{Table IV: Comparison of robustness performance of Burger and other baselines.}
\label{robust analysis}
\resizebox{1\linewidth}{!}{
\begin{tabular}{ccccccc}
\specialrule{1pt}{0pt}{0pt}
\multirow{2}{*}{Models} & \multicolumn{2}{c}{Ciao(30\% Synthetic   Data)} & \multicolumn{2}{c}{Douban(30\% Synthetic   Data)} & \multicolumn{2}{c}{Yelp(30\% Synthetic Data)} \\
                        & HR@3                   & NDCG@3                 & HR@3                    & NDCG@3                  & HR@3                  & NDCG@3                \\ \hline
SMIN                    & 0.4873                 & 0.4082                 & 0.6117                  & 0.4988                  & 0.3728                & 0.2891                \\
Dec\%                   & 4.15\%                 & 4.51\%                 & 3.30\%                  & 4.21\%                  & 4.34\%                & 5.77\%                \\ \hline
DSL                     & 0.5033                 & 0.4286                 & 0.6169                  & 0.5048                  & 0.3901                & 0.3098                \\
Dec\%                   & 2.31\%                 & 1.27\%                 & 2.00\%                  & 2.53\%                  & 3.13\%                & 3.07\%                \\ \hline
Burger                  & 0.5174                 & 0.4472                 & 0.6512                  & 0.5473                  & 0.3991                & 0.3165                \\
Dec\%                   & 3.58\%                 & 3.43\%                 & 3.57\%                  & 3.01\%                  & 4.20\%                & 4.38\%                \\ 
\specialrule{1pt}{0pt}{0pt}
\end{tabular}
}
\end{table}
\section{Conclusion}
In this paper, we propose a social recommendation model named Burger. It enhances recommendation performance by modeling multi-semantic influences and fusing graph denoising and augmentation. Burger constructs a social tensor and updates it via a sliding window. It jointly mines user item and social preferences, uses a bi-semantic loss for multi-semantic modeling, and leverages Bayesian probability to denoise and find potential social relations. \par
In the future, we plan to explore the use of large language models for purifying social data. We'll test different types of large language models and their unique capabilities in handling social data, aiming to find the most effective approach for enhancing data quality.
\section*{Acknowledgement}
This work is supported by Beijing Advanced Innovation Center for Future Blockchain and Privacy Computing.
\bibliographystyle{IEEEbib}
\bibliography{ref}
\end{document}